\documentclass[twocolumn,prl,showpacs,superscriptaddress,floatfix]{revtex4-1}
\usepackage{graphicx}
\usepackage{amsmath}
\usepackage{amssymb}
\usepackage{color}
\usepackage{bm}

\newcommand{\bmx}{{\bm{x}}}
\newcommand{\bmv}{{\bm{v}}}
\newcommand{\bmF}{{\bm{F}}}
\newcommand{\bmxi}{{\bm{\xi}}}
\newcommand{\bmvhat}{\hat{\bm{v}}}

\begin{document}
\title{Macroscopic Time-Reversal Symmetry Breaking at Nonequilibrium Phase
Transition}

\author{Pyoung-Seop Shim}
\affiliation{Department of Physics, University of Seoul, Seoul 130-743,
Korea}
\author{Hyun-Myung Chun}
\affiliation{Department of Physics, University of Seoul, Seoul 130-743,
Korea}
\author{Jae Dong Noh}
\affiliation{Department of Physics, University of Seoul, Seoul 130-743,
Korea}
\affiliation{School of Physics, Korea Institute for Advanced Study,
Seoul 130-722, Korea}
\date{\today}

\begin{abstract}
We study the entropy production in a macroscopic nonequilibrium system that
undergoes an order-disorder phase transition. 
Entropy production is a characteristic feature of nonequilibrium dynamics
with broken detailed balance. 
It is found that the entropy production rate per particle vanishes
in the disordered phase and becomes positive in the ordered phase following
critical scaling laws.
We derive the scaling relations for associated critical exponents. 
Our study reveals that a nonequilibrium ordered state is sustained at the
expense of macroscopic time-reversal symmetry breaking with an extensive
entropy production while a disordered state costs only a subextensive
entropy production.
\end{abstract}


\maketitle

Detailed balance is the hallmark of the thermal equilibrium state. A system
is said to obey detailed balance if the probability current along any
microscopic trajectory in the phase space is balanced by that 
along the time-reversed one~\cite{{Gardiner:2010tp}}. 
Consequently, time-reversal symmetry is preserved in thermal 
equilibrium.

Thermodynamics of nonequilibrium systems, where detailed balance and
time-reversal symmetry are broken with a positive entropy production, 
has been attracting a lot of 
interests~\cite{{Evans:1993ix},{Gallavotti:1995wv},{Jarzynski:1997uj},
{Sekimoto:1998uf},{Lebowitz:1999tv},{Crooks:1999ta},{Seifert:2005fu}}. 
Recent studies have been focused on microscopic systems with a few degrees 
of freedom where the effect of thermal fluctuations are strong.
Under the framework of stochastic thermodynamics, various 
fluctuation theorems are discovered, which provide useful insights 
on the nature of nonequilibrium fluctuations. Theoretical works
foster experimental studies of microscopic systems such as 
molecular motors, nano heat engines, biomolecules, 
and so on~\cite{{Wang:2002hw},{Carberry:2004va},{Collin:2005wp},{Blickle:2011bm},{GomezSolano:2011em},{Lee:2015hn}}. 

Macroscopic systems pose an intriguing question on the level of 
irreversibility. Consider a many-particle system displaying an order-disorder
phase transition whose microscopic dynamics does not obey detailed balance.
Does the broken detailed balance result in time-reversal symmetry breaking 
at the macroscopic level?
On the one hand, one may expect that entropy productions of each
particle add up to a macroscopic amount irrespective of a macroscopic state.
On the other hand, if the system is in a disordered phase 
so that all configurations are almost equally likely, 
then irreversibility may not show up on a macroscopic level producing 
only a subextensive amount of entropy.
A system in an ordered state has a lower entropy than in a disordered state.
Then, which phase produces more total entropy including the system entropy and
the environmental entropy?
These questions lead us to the study of the entropy production in a
model system undergoing nonequilibrium phase transition. 

In this paper, we investigate 
the emergence of macroscopic irreversibility  
out of microscopic dynamics with broken detailed balance.
We find that the total entropy production changes its character from being 
subextensive to being extensive as the system undergoes an order-disorder
phase transition. The entropy production rate per particle exhibits 
critical scaling laws as an order parameter does in ordinary critical
phenomena, and scaling
relations among critical exponents are derived. 
Although the results are derived in a specific model system, we argue that
the scaling behaviors should be valid for general nonequilibrium systems.

As a nonequilibrium model, we adopt the particle system in two
dimensions introduced in Ref.~\cite{{Sevilla:2014bx}}. 
This model describes a flocking phenomenon of passive particles.
In nature a flock of birds and a school of fish
display a collective motion~\cite{Vicsek:2012gp}. 
Such a phenomenon has been studied with 
microscopic models consisting of active self-propelled 
particles moving at a constant speed~\cite{{Vicsek:1995ti},{Gregoire:2004uf}}. 
Flocking takes place when particles are subject to an 
interaction that favors alignment of individual velocities to the 
mean direction.

The model in Ref.~\cite{Sevilla:2014bx} is composed of passive particles
in the thermal reservoir instead of active particles.
It consists of $N$ Brownian particles of mass $m$ 
in a two-dimensional plane of size $L\times L$ embedded 
in a thermal reservoir at constant temperature $T$. 
The particle density is denoted by $\rho=N/L^2$.  
Let $\bmx_i=(x_{i1},x_{i2})$ and $\bm{v}_i = \frac{d\bmx_i}{dt} =(v_{i1},v_{i2})$ 
be the position and the velocity of a particle $i=1,\cdots,N$.
We will represent a configuration of the whole system with a short-hand
notation $\bm{Z} = (\bm{X},\bm{V})$ with $\bm{X} = \{\bmx_1,\bmx_2,\cdots,\bmx_N\}$ and
$\bm{V} = \{\bmv_1,\bmv_2,\cdots,\bmv_N\}$.
The equations of motion are given by
\begin{equation}\label{eq:Lang}
m\frac{d\bmv_i}{dt} = \bmF_i(\bm{V}) -\gamma \bmv_i + \bmxi_i(t)
\end{equation}
where $\gamma$ is the damping coefficient and $\bmxi_i(t) =
(\xi_{i1}(t),\xi_{i2}(t))$ is the thermal noise satisfying
\begin{equation}\label{eq:noise.correl}
\begin{aligned}
\left \langle \xi_{ia}(t) \right \rangle &=0\\
\left \langle \xi_{ia} (t) \xi_{jb}(t') \right \rangle &=
2\gamma k_B T \delta_{ij}\delta_{ab} \delta(t-t')
\end{aligned}
\end{equation}
with the Boltzmann constant $k_B$, which will be set to unity hereafter.
The velocity aligning force $\bmF_i(\bm{V})$ is taken to be 
\begin{equation}\label{Fi_def}
\bmF_i(\bm{V}) = \Gamma \bmvhat_i \times (\bm{f} \times \bmvhat_i)  =
\Gamma \left( \bm{f} - (\bm{f} \cdot \bmvhat_i ) 
\bmvhat_i \right), 
\end{equation}
where $\Gamma$ is the interaction strength, $\bmvhat_i = \bmv_i / |\bmv_i|$
is the unit vector, and 
\begin{equation}\label{eq:meandirec}
\bm{f}=\frac{1}{N}  \sum _{j=1}^{N}\bmvhat_j \ .
\end{equation}
The vector $\bm{f}$ points towards the average direction of the
particles, and its magnitude $\Lambda = |\bm{f}|$ plays a role of the order
parameter for the collective motion. 
Note that the force $\bm{F}_i$ is perpendicular to $\bmv_i$. 
It does not work on the particle but turns the direction of $\bmv_i$ toward
$\bm{f}$.
The interaction is infinite-ranged. 
A short-ranged version of the model was studied in 
Ref.~\cite{Dossetti:2014vk}.

Numerical study in Ref.~\cite{Sevilla:2014bx} found that 
the system undergoes a phase transition separating a disordered
phase~($\Gamma<\Gamma_c$) and an ordered phase~($\Gamma>\Gamma_c$).
Near $\Gamma = \Gamma_c$, 
the order parameter scales as $\langle \Lambda \rangle_s
\sim (\Gamma-\Gamma_c)^\beta$ and the susceptibility $\chi \equiv N (\langle
\Lambda^2\rangle_s - \langle \Lambda \rangle_s^2 )$ scales as $\chi \sim
|\Gamma-\Gamma_c|^{-\gamma}$, 
where $\langle \rangle_s$ denotes the steady-state ensemble average. 
The critical exponents are given by
$\beta/\nu \simeq 0.491$ and $\gamma/\nu \simeq 1.02$, where $\nu \simeq
0.94$ is the correlation length 
exponent~($\xi\sim |\Gamma-\Gamma_c|^{-\nu}$)~\cite{{Sevilla:2014bx}}. 
These exponents are compatible with those of the mean field XY
model~\cite{{Kim:2001jt},{Sasa:2015kf}}.
When the interaction is infinite-ranged, the correlation volume
$\xi_V$ is more useful than the correlation length $\xi$.
Since the model under consideration is embedded in the two-dimensional
space, the correlation volume is given by $\xi_V = \xi^2$ and scales as
$\xi_V \sim |\Gamma-\Gamma_c|^{-\bar{\nu}}$ with $\bar{\nu} = 2\nu$. 

The velocity-dependent force breaks the detailed balance and 
the time-reversal symmetry.
We quantify the amount of the time-reversal symmetry breaking by
the entropy production. Suppose that the system
evolves along a stochastic trajectory $\mathcal{Z}[\tau] =
\{(\bm{X}(t),\bm{V}(t))|0\leq t \leq \tau\}$ for a time 
interval $\tau$. Following stochastic thermodynamics~\cite{Seifert:2005fu},
the total entropy production $\Delta S_{\rm tot}[\mathcal{Z}[\tau]]$ 
along a given trajectory $\mathcal{Z}[\tau]$ is determined by 
the probability ratio of $\mathcal{Z}[\tau]$ against its time-reversed
trajectory $\mathcal{Z}^R[\tau] = \{ (\bm{X}(\tau-t),-\bm{V}(\tau-t))| 0\leq
t\leq
\tau\}$~\cite{{Seifert:2005fu},{Spinney:2012di},{Lee:2013fb},{Ganguly:2013db},{Chaudhuri:2014bg},{Kwon:2015vj}}. 

In our model, the total entropy production is decomposed into three terms 
as~\cite{note_app}
\begin{equation}
\Delta S_{\rm tot}[\mathcal{Z}] = \Delta S_{\rm sys}[\mathcal{Z}] -
\frac{Q[\mathcal{Z}]}{T} + \Delta S_{\rm v}[\mathcal{Z}] , 
\end{equation}
where $\Delta S_{\rm sys}$ is the change in the Shannon entropy of the
system, the second term is the Clausius form for the entropy change 
of the heat bath with $Q$ being the heat absorbed by the system,
and the last term $S_{\rm v}$ appears only in the presence of a
velocity-dependent force~\cite{Kwon:2015vj} and is given by
\begin{equation}\label{S_v}
\Delta S_{\rm v}[\mathcal{Z}] = \frac{m}{\gamma T}\sum_{i=1}^N \int_0^\tau dt
\bm{F}_i(\bm{V}(t)) \circ \frac{d\bmv_i(t)}{dt} .
\end{equation}

In the steady state, the ensemble average of $\Delta S_{\rm sys}$
vanishes. The thermodynamic first law reads as $\Delta E = Q + W$
where $\Delta E$ is the change in the total energy 
$E = \sum_i \frac{1}{2} m\bmv_i^2$ and $W = \sum_i \int_0^t dt 
\bmv_i \cdot \bmF_i$ is the work done by the force.
Since $W=0$, $Q = \Delta E$ and its steady state average vanishes. 
Thus, the entropy production rate per particle in the steady state is given by
\begin{equation}
s \equiv \frac{1}{N} \left\langle \frac{dS_{\rm tot}}{dt}\right\rangle_s = \frac{m}{\gamma T} \frac{1}{N} \sum_{i=1}^N \left\langle \bmF_i \circ
\frac{d\bmv_i}{dt} \right\rangle_s .
\end{equation}

\begin{figure}
\includegraphics*[width=\columnwidth]{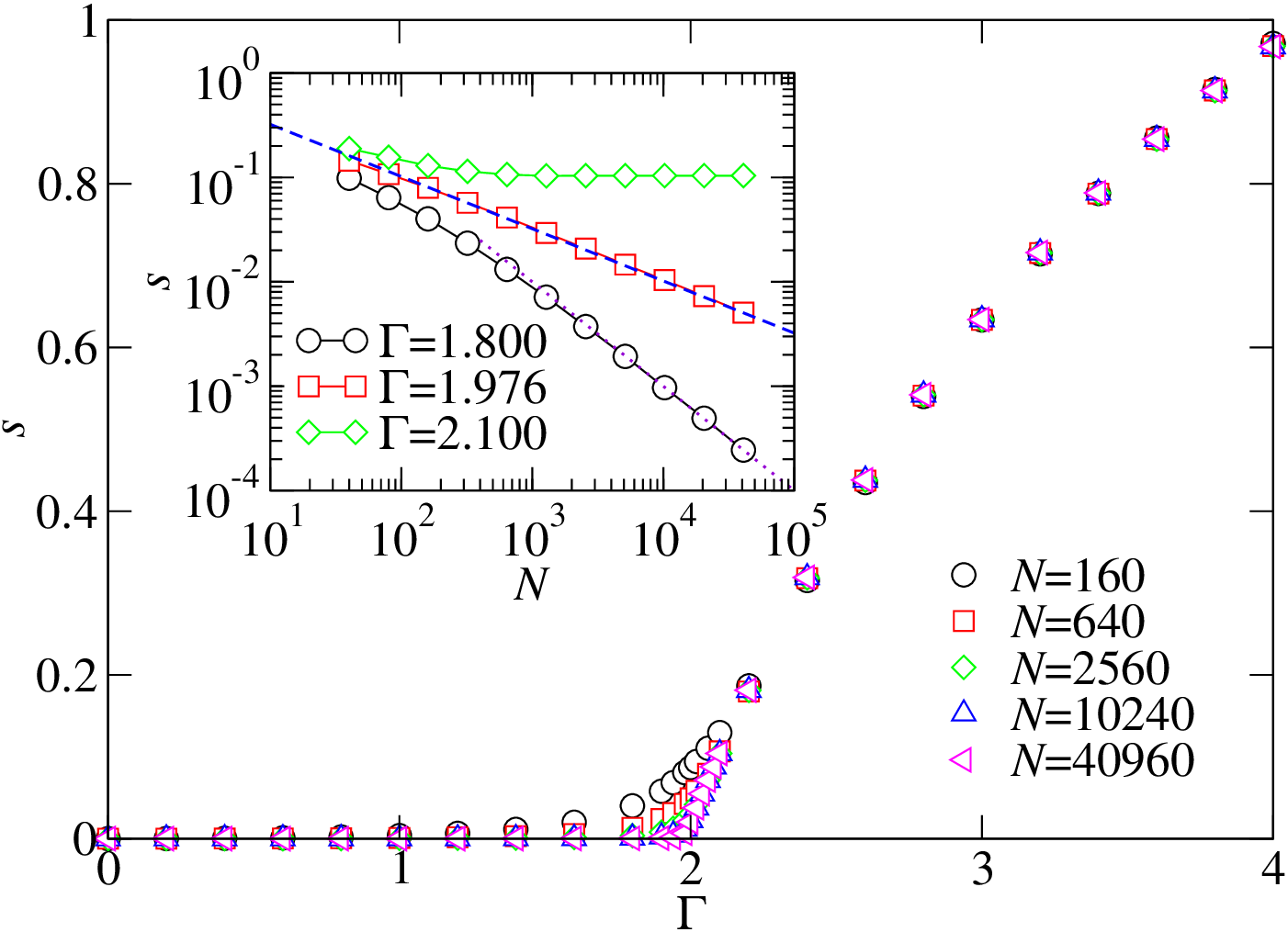}
\caption{
$s$ versus $\Gamma$ for several values of $N$. Inset shows the finite size
scaling behaviors of $s$ when $\Gamma$ is below, equal to, and above
$\Gamma_c=1.976$. The dotted~(dashed) line has the slope $-1$~($-1/2$).
}\label{fig:entropy}
\end{figure}  

\begin{figure}
\includegraphics*[width=\columnwidth]{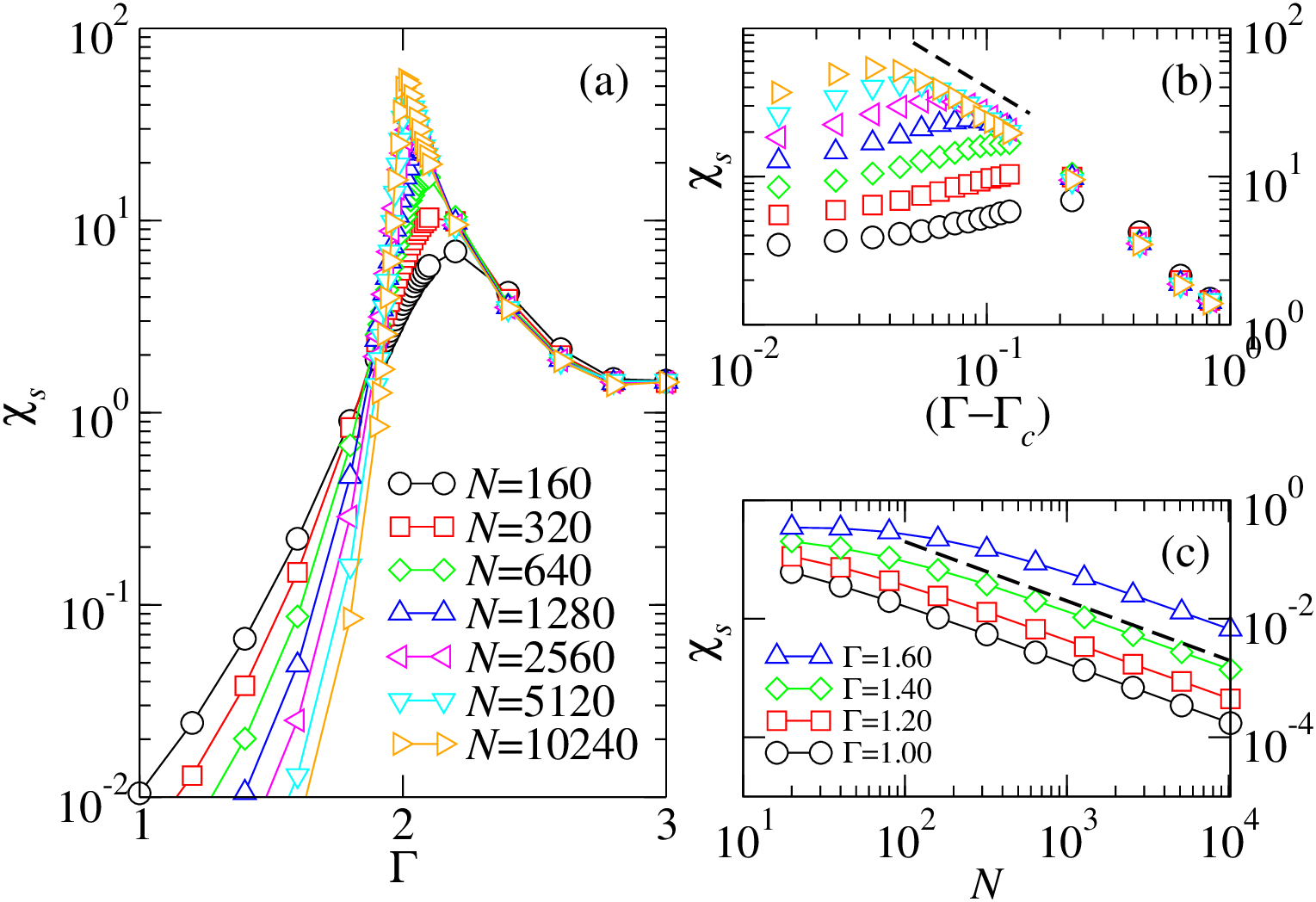}
\caption{ (a) Susceptibility $\chi_s$ as a function of $\Gamma$.
(b) $\chi_s$ versus $(\Gamma-\Gamma_c)$ in the log-log scale. The dashed
line has the slope $-1$.
(c) $\chi_s$ versus $N$ for $\Gamma < \Gamma_c$. The dashed line has the
slope $-1$.}
\label{fig:suscep}
\end{figure}

We have performed numerical simulations. 
The equations of motion in \eqref{eq:Lang} 
are integrated numerically by using the time-discretized~($\Delta t = 0.01$) 
Heun algorithm~\cite{Greiner:1988cq}.
We took $m=\gamma=\rho=2T = 1$ in all simulations.
Figure~\ref{fig:entropy} shows that 
$s$ displays a characteristic behavior signaling 
a continuous phase transition. 
As $N$ increases, $s \sim 1/N$ for $\Gamma < \Gamma_c$ while
it converges to a finite value for $\Gamma> \Gamma_c$. 
We also measure the susceptibility of the entropy production that is
defined as
\begin{equation}\label{chi_s}
\chi_s(\Gamma,N,\tau) = \frac{1}{\tau N} \left[ \left\langle \Delta S_{\rm
v}^2
\right \rangle_s - \left\langle \Delta S_{\rm v}\right\rangle_s^2\right] ,
\end{equation}
where $\Delta S_{\rm v}$ denotes the entropy production of $N$ particles 
in a time interval $\tau$. Figure~\ref{fig:suscep} (a) shows the
susceptibility measured at fixed $\tau=64$. It has a sharp peak at
$\Gamma=\Gamma_c$, which also reminds us of a continuous phase transition.
The threshold $\Gamma_c \simeq 1.976$ is close to the onset of the
collective motion reported in Ref.~\cite{Sevilla:2014bx}. 
We will show that the entropy production indeed exhibits the continuous
phase transition and that the phase transition is triggered by the 
onset of the collection motion.

The entropy production can be related to the order parameter $\Lambda$ 
for the collective motion. Using the equations of motion for $d\bmv_i/dt$, 
the entropy production in
\eqref{S_v} is written as~\cite{note_app}
\begin{equation}
\begin{aligned}
\Delta S_{\rm v} = & \sum_{i=1}^N \int_0^\tau dt \left[ \frac{1}{\gamma T}
|\bmF_i|^2 + \frac{1}{m} \bm{\nabla}_{\bmv_i} \cdot \bmF_i \right]\  \\
   & + \sum_{i=1}^N \frac{1}{\gamma T} \int_0^\tau \bmF_i \cdot
d\bm{W}_i(t) ,
\end{aligned}
\end{equation}
where $\bm{\nabla}_{\bmv_i}$ denotes the gradient operator with
respective to $\bmv_i$ and $d\bm{W}_i(t) = \int_t^{t+dt}dt' \bmxi_i(t')$.
The last term contributes neither to the ensemble average nor to the
susceptibility because it is of the order of $\mathcal{O}(\tau^{1/2})$ 
with zero mean while the others scale linearly with $\tau$. 
Hence, it will be ignored.
We then introduce the polar coordinate so that 
the velocity vector is written as
$\bmv_i = (v_i \cos\theta_i,v_i \sin\theta_i)$. The relation
\eqref{eq:meandirec} for the vector 
$\bm{f} = (\Lambda \cos\psi,\Lambda \sin\psi)$ is written as
\begin{equation}\label{Lambda_psi}
\Lambda e^{i\psi} = \frac{1}{N} \sum_{j=1}^N e^{i\theta_j} \ .
\end{equation}
By using \eqref{Fi_def} and \eqref{Lambda_psi}, we can show
that~\cite{note_app}
\begin{equation}\label{Sm_polar}
\Delta S_{\rm v} = \sum_{i=1}^N \int_0^\tau dt \left[ A_i - B_i + C_i\right] +
\mathcal{O}(\tau^{1/2}),
\end{equation}
where $A_i = \frac{\Gamma^2\Lambda^2}{\gamma T} \sin^2(\psi-\theta_i)$,
$B_i=\frac{\Gamma\Lambda}{m v_i}\cos(\psi-\theta_i)$, and $C_i =
\frac{\Gamma}{Nmv_i}$.

The expression in \eqref{Sm_polar} gives a hint on the scaling behavior of 
the entropy production.
The macroscopic variables $\Lambda$ and $\psi$ fluctuate much slower than 
the microscopic variables $v_i$'s and $\theta_i$'s. Thus, in taking
the ensemble-average of \eqref{Sm_polar}, we can use the adiabatic
approximation~\cite{Sasa:2015kf} to replace $\Lambda^2$ and
$\Lambda$ with their ensemble averaged values. Power counting combined with
the adiabatic approximation leads to the conclusion that
the entropy production rate per particle scales as
$s \sim {\langle \Lambda^2\rangle_s} \sim \langle \Lambda\rangle_s^2$~(from
$A_i$ and $B_i$) with the $\mathcal{O}(N^{-1})$ correction~(from $C_i$).
Therefore, we expect that the entropy production rate per particle exhibits 
a critical power law scaling
\begin{equation}\label{s_power_law}
s  \sim (\Gamma-\Gamma_c)^{\beta_e}
\end{equation}
with the critical exponent
\begin{equation}\label{scaling_rel_s}
\beta_e = 2 \beta 
\end{equation}
for $\Gamma>\Gamma_c$ and $s \sim 1/N$ for $\Gamma<\Gamma_c$.
When $N$ is finite, following the standard finite-size-scaling~(FSS) ansatz,
we expect that
\begin{equation}\label{s_FSS}
s = N^{-\beta_e / \bar{\nu}}
\Phi\left((\Gamma-\Gamma_c)N^{1/\bar{\nu}}\right) .
\end{equation}
The scaling function $\Phi(x)$ has the limiting behaviors $\Phi(x)
\stackrel{x\gg 1}{\longrightarrow} x^{\beta_e}$ ensuring
\eqref{s_power_law} and $\Phi(x) \stackrel{x\ll
-1}{\longrightarrow} |x|^{\beta_e-\bar{\nu}}$ guaranteeing the 
$N^{-1}$ scaling in the disordered phase.

\begin{figure}
\includegraphics*[width=\columnwidth]{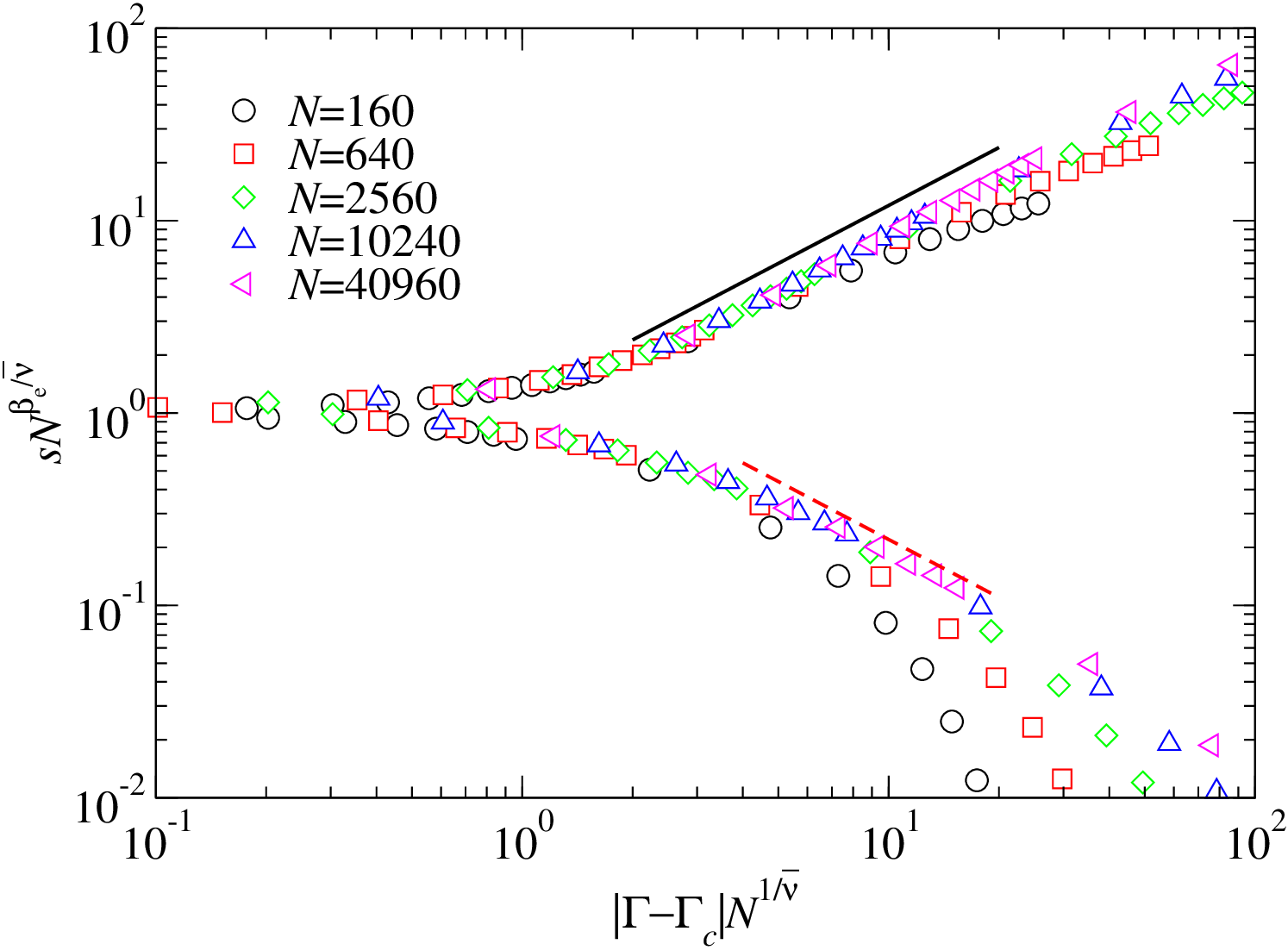}
\caption{(Color online) Scaling plot of $s N^{\beta_e/\bar\nu}$ versus
$|\Gamma-\Gamma_c| N^{1/\bar\nu}$ according to \eqref{s_FSS}.
The solid~(dashed) line has slope $1~(-1)$.} 
\label{fig:fss}
\end{figure}

The numerical data in Fig.~\ref{fig:entropy} are analyzed according to the
FSS form with the mean field critical exponents $\beta_e=1$ and $\bar\nu=2$. As shown
in Fig.~\ref{fig:fss}, the data collapse and the limiting behaviors 
of the scaling function confirm the scaling relation in 
\eqref{scaling_rel_s} and the FSS form of \eqref{s_FSS}.

The total entropy production $\Delta S_{\rm v}$ is given by the spatial and
temporal sum of the fluctuating local entropy production rates. We can
derive the scaling form for the susceptibility $\chi_s$ in the following
way: Near the critical point, the correlation volume and time diverge as 
$\xi_V \sim |\Gamma-\Gamma_c|^{-\bar\nu}$ and $\xi_t \sim
|\Gamma-\Gamma_c|^{-\nu_t}$, respectively.
When $N\gg \xi_V$ and $\tau \gg \xi_t$ in the ordered 
phase~($\Gamma>\Gamma_c$), the total entropy production $\Delta S_{\rm v}$
is the sum of the contributions from $M=\tau N/(\xi_t\xi_V)$ space-time blocks. 
All the blocks are independent because they are beyond the correlation volume 
and time. Therefore, the susceptibility should scale as 
$\chi_s \sim \frac{1}{\tau N} \times M \times (\xi_t\xi_V s)^2 \sim
\xi_t\xi_V s^2$, which leads to the scaling form 
\begin{equation}\label{chi_s_gamma_e}
\chi_s \sim (\Gamma-\Gamma_c)^{-\gamma_e}
\end{equation}
with the susceptibility exponent
\begin{equation}\label{hyperscaling}
\gamma_e = \nu_t + \bar\nu - 2 \beta_e .
\end{equation}
This is the hyperscaling relation extended to the systems with anisotropic
scaling~\cite{Hong:2007di,Henkel:2001hy}.
At the critical point, the finite-size effect dominates so that
\begin{equation}\label{chi_FSS}
\chi_s(\Gamma_c,N,\tau) \sim \left\{
\begin{array}{ccc} 
              \tau^{\gamma_e/\nu_t} &,& \tau \ll N^{\bar{z}} \\ [2mm]
              N^{\gamma_e/\bar\nu} &,& \tau \gg N^{\bar{z}} 
\end{array}\right.
\end{equation}
with $\bar{z} = \nu_t/\bar\nu$. In the disordered phase, the entropy
production rate per particle vanishes as $s\sim 1/N$, so does the
susceptibility $\chi_s \sim 1/N$. 

\begin{figure}
\includegraphics*[width=\columnwidth]{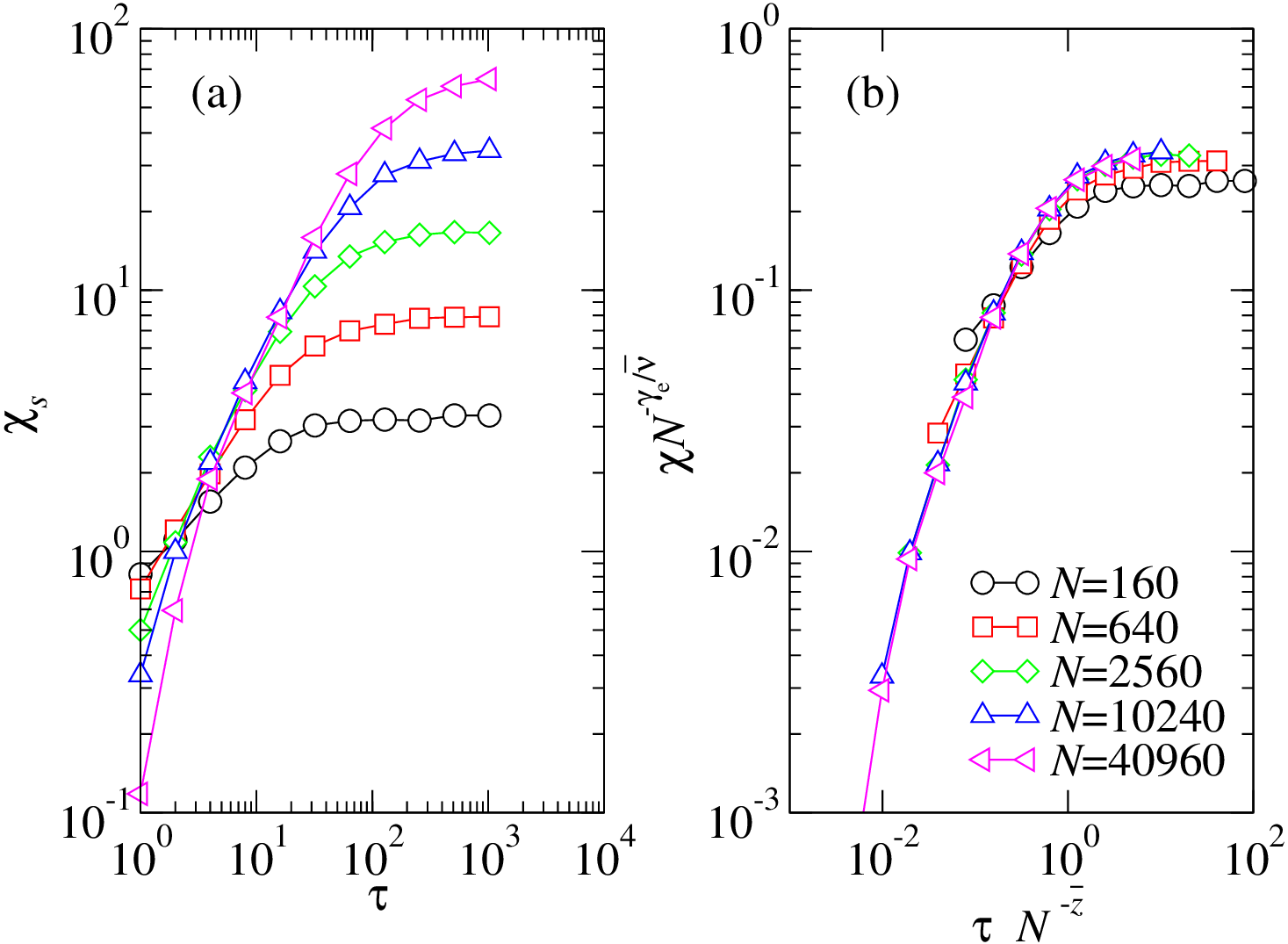}
\caption{(Color online)
(a) $\chi_s(\Gamma_c,N,\tau)$ versus $\tau$ at several values of $N$.
(b) Scaling plot of $\chi_s(\Gamma_c,N,\tau) N^{-\gamma_e/\bar\nu}$ versus
$\tau N^{-\bar{z}}$.
}
\label{fig:chiFSS}
\end{figure}

The numerical data support the scaling theory.
Figure~\ref{fig:suscep} (b) shows the susceptibility follows
the power law of \eqref{chi_s_gamma_e} with $\gamma_e = 1$. 
This exponent value satisfies the hyperscaling
relation in \eqref{hyperscaling} with $\nu_t=1$, $\bar\nu=2$, and
$\beta_e=1$. The $1/N$ scaling inside the disordered phase is also
checked in Figure~\ref{fig:suscep} (c). 
The FSS behavior at the critical point
$\Gamma=\Gamma_c$ is examined in Fig.~\ref{fig:chiFSS}.
At a given $N$, $\chi_s(\Gamma_c,N,\tau)$ 
increases algebraically with $\tau$ and saturates to a limiting value~(see
Fig.~\ref{fig:chiFSS}(a)). The scaling plot in Fig.~\ref{fig:chiFSS}(b)
confirms the scaling behavior of \eqref{chi_FSS} for $\tau\ll N^{\bar{z}}$
and $\tau\gg N^{\bar{z}}$.

We have shown that the broken detailed balance leads to the macroscopic 
entropy production only in the ordered phase
using the analytic scaling theory and the numerical simulations.
The entropy production per particle per unit time $s$  is positive 
but vanishes as $1/N$ in the disordered phase, while it is finite 
in the ordered phase following the power law~(see \eqref{s_power_law}). 
The susceptibility $\chi_s$ vanishes in the disordered phase and 
follows the power law~(see \eqref{chi_s_gamma_e}) in the ordered phase. 
The critical exponents satisfy the scaling relations in 
\eqref{scaling_rel_s} and \eqref{hyperscaling}.

The quadratic relation $s\sim \langle \Lambda\rangle_s^2$ is crucial in
deriving the scaling theory. This relation is derived in a model
system that has a mean field nature. 
We argue that the scaling behaviors are universal in general 
thermal systems undergoing a nonequilibrium phase transition between a
disordered phase and an ordered phase. Collective motions in the ordered
phase are characterized by the thermodynamic currents $J_i$ of e.g.,
energy and particle. The currents are small near the critical point. 
Thus, following the linear irreversible thermodynamics of 
Onsager~\cite{Onsager:1931uu}, one can assume that 
$J_i = \sum_j L_{ij} X_j$ where $X_j$'s are the thermodynamic forces
and $L_{ij}$'s are the Onsager coefficients. The entropy production rate is
then given by $dS/dt = \sum_i X_i J_i = \sum_{i,j} L^{-1}_{ij} J_i
J_j\propto J^2$, which supports the validity of the quadratic relation
between the entropy production rate and the current density. 
In stochastic thermodynamics, the total entropy production rate is written 
as the configuration space average of the probability current density
squared~\cite{Seifert:2005fu}, which also supports the relation. It would be
interesting to investigate the scaling relations in \eqref{scaling_rel_s} 
and \eqref{hyperscaling} in systems with a short-ranged interaction.

The result that the ordered phase costs more environmental
entropy production may be understood in the framework of the thermodynamic 
second law. Suppose that one changes a coupling
constant of a system so that it relaxes from a disordered phase to an ordered
phase in a characteristic relaxation time $t_{\rm relax}$. 
During the process, the system
entropy decreases at the rate $dS_{\rm sys}/dt \sim 
\Delta S_{\rm sys}/t_{\rm relax} =
(S_{\rm sys}({\rm ordered})-S_{\rm sys}({\rm disordered}))/t_{\rm relax} < 0$.
The thermodynamic second law requires that the entropy production rate
should be nonnegative at any moment. 
Therefore, during the relaxation process, 
the environmental entropy production rate should satisfy $d{S}_{\rm env}/dt
\geq - dS_{\rm sys}/dt \sim | \Delta S_{\rm sys}|/t_{\rm relax}$, 
which gives a lower bound for the environmental entropy production rate.
It should be investigated further whether the inequality is working in the
steady state. We leave it for future work.

This work was supported by the Basic Science Research Program through the
NRF Grant No. 2013R1A2A2A05006776.

\appendix
\bibliographystyle{apsrev}
\bibliography{preprint}

\begin{widetext}
\section{Appendix: Total entropy production}

It is straightforward to decide whether a deterministic dynamics 
is reversible or not.
Suppose that a system evolves from a configuration $Z(0)=(X(0),V(0))$ to
$Z(\tau)=(X(\tau),V(\tau))$ along a trajectory 
$\mathcal{Z}[\tau]=\{Z(t)| 0\leq t\leq \tau\}$. 
If one flips the velocity in the final configuration and takes the resulting
configuration $(X(\tau),-V(\tau))$ as an initial state, then the reversible 
dynamics lets the system follow the time-reversed trajectory 
$\mathcal{Z}^R[\tau] = \{ Z^R(t) | 0
\leq t \leq \tau\}$ with $Z^R(t) \equiv (X(\tau-t),-V(\tau-t))$.

Generalizing this idea to stochastic systems, one can define 
the irreversibility or the entropy production by comparing the probability 
of trajectories $\mathcal{Z}[\tau]$ and $\mathcal{Z}^R[\tau]$. 
The probability distribution function~(PDF) of a given trajectory 
$\mathcal{Z}[\tau]$ is given by
$P[\mathcal{Z}[\tau]] = \Pi\left[\mathcal{Z}[\tau] ; Z(0)\right]\ p_0
(Z(0))$, where $p_0(Z)$ is an initial PDF of being in a configuration 
$Z$ at time $t=0$ and $\Pi[\mathcal{Z}[\tau] ; Z(0)]$ is 
a conditional probability distribution of $\mathcal{Z}[\tau]$ 
to a given initial configuration $Z(0)$. The PDF for a
time-reversed trajectory $\mathcal{Z}^R[\tau]$
is similarly given by 
$P[\mathcal{Z}^R[\tau]] = \Pi\left[\mathcal{Z}^R[\tau] ; Z^R(0)\right]\ 
p_\tau (Z(\tau))$,
where $p_\tau(Z)$ is the PDF at time $\tau$ 
which has evolved from $p_0(Z)$.
According to stochastic thermodynamics, 
the total entropy production for a given
trajectory $\mathcal{Z}[\tau]$ is given by~\cite{Seifert:2005fu}
\begin{equation}\label{eq:definition}
\Delta S_{\text{tot}}  = \ln 
  \frac{\Pi[\mathcal{Z}[\tau];Z(0)]\ p_0(Z(0))}
       {\Pi[\mathcal{Z}^R[\tau];Z^R(0)]\ p_\tau(Z(\tau))} \ .
\end{equation}
It consists of two parts as $\Delta S_{\text{tot}} = \Delta S_{\text{sys}} +
\Delta S_{\text{env}}$, where 
\begin{equation}
\Delta S_{\text{sys}} = -\ln p_\tau(Z(\tau)) + \ln p_0(Z(0))
\end{equation}
is the system entropy change and the remaining term 
$\Delta S_{\text{env}}$ is the environmental entropy production. 

The environmental entropy production can be written in terms of physical
quantities. This task has been done in a recent preprint~\cite{Kwon:2015vj} for systems with an
arbitrary velocity-dependent force. We make use of Eq.~(11) of 
Ref.~\cite{Kwon:2015vj} to obtain that
\begin{equation}
\Delta S_{\text{env}} = -\frac{m}{T} \sum_{i=1}^N \int_0^\tau 
\bm{v}_i(t) \circ d\bm{v}_i(t) +
\frac{m}{\gamma T} \sum_{i=1}^N \int_0^\tau \bm{F}_i(\bm{V}(t)) \circ
d\bm{v}_i(t) ,
\end{equation}
where the notation $\bm{A}(t) \circ d\bm{B}(t) \equiv 
\frac{\bm{A}(t+dt)+\bm{A}(t)}{2} \cdot (\bm{B}(t+dt)-\bm{B}(t))$ 
stands for the stochastic integral in the Stratonovich
sense~\cite{Gardiner:2010tp}. 
Using $m d\bm{v}_i = \left( m d\bm{v}_i - \bm{F}_i dt \right) + 
\bm{F}_i dt$ in the first term, one can further decompose $\Delta
S_{\text{env}}$ as
\begin{equation}
\Delta S_{\text{env}} = -\frac{ \sum_{i=1}^N \int_0^\tau \bm{v}_i 
\circ (m d\bm{v}_i -
\bm{F}_i dt)}{T} - \frac{\sum_{i=1}^N \int_0^\tau \bm{v}_i \circ 
\bm{F}_i dt }{T} +
\frac{m}{\gamma T} \sum_{i=1}^N \int_0^\tau \bm{F}_i(\bm{V}) \circ d\bm{v}_i \ .
\end{equation}
The Langevin equation indicates that 
\begin{equation}
Q = \sum_{i=1}^N \int_0^\tau \bm{v}_i \circ \left(m \frac{d\bm{v}_i}{dt} -
\bm{F}_i\right) dt
= \sum_{i=1}^N \int_0^\tau \bm{v}_i \circ  
\left( -\gamma \bm{v}_i + \bm{\xi}_i \right)dt
\end{equation} 
is the work done by the heat bath through the damping force 
and the random force, namely the heat absorbed by the system from the heat
bath. The second term is identically zero since $\bm{v}_i \perp
\bm{F}_i$. The third term is $\Delta S_{\rm v}$. 
This completes the derivation of Eqs.~(5) and (6) of the main text. 
In $\Delta S_{\rm tot}$, $(\Delta S_{\rm sys} - Q/T)$ is generic in all thermal
systems, while the others appear only in the presence of velocity-dependent
forces.

Note that the force $\bm{F}_i$ does not work~($W=0$). 
Consequently, the thermodynamic first law is written
as $\Delta E = Q$,
where $\Delta E$ is the change in the total kinetic energy 
$E \equiv \sum_{i=1}^N \frac{1}{2}m\bm{v}_i^2$.

\section{Appendix: Derivation of Eqs.~(9) and (11)}
The Stratonovich product $\bm{F}_i \circ d\bm{v}_i$ is defined
as~\cite{Gardiner:2010tp}
\begin{equation}\label{sp}
\bm{F}_i \circ d\bm{v}_i =
\frac{\bm{F}_i(\bm{V}(t+dt))+\bm{F}_i(\bm{V}(t))}{2} \cdot d\bm{v}_i(t)
= \sum_{a} F_{ia}(t) dv_{ia} + \frac{1}{2} \sum_{j,a,b} \frac{\partial
F_{ia}}{\partial v_{jb}} dv_{ia} dv_{jb} + o(dt) \ ,
\end{equation}
where $i,j=1,\cdots,N$ are particle indices and $a,b=1,2$ are Cartesian
coordinate indices.
We now use the Langevin equation to replace $md\bm{v}_i = \bm{F}_i dt -
\gamma \bm{v}_i dt + d\bm{W}_i$, where $d\bm{W}_i = \int_t^{t+dt} dt'
\bm{\xi}_i(t')$ satisfying that $\langle dW_{ia}\rangle = 0$ and $\langle
dW_{ia} dW_{jb}\rangle = 2\gamma T \delta_{ij}\delta_{ab}dt$. 
Inserting this into \eqref{sp}, we obtain that 
\begin{equation}
m\bm{F}_i \circ d\bm{v}_i = |\bm{F}_i|^2 dt + \bm{F}_i \cdot d\bm{W}_i
+\frac{1}{2m} \sum_{j,a,b} \frac{\partial F_{ia}}{\partial v_{jb}}
dW_{ia}dW_{jb} + o(dt) .
\end{equation}
Since $dW_{ia}$'s are independent of each other, one can replace
$(dW_{ia}dW_{jb})$  with $(2\gamma
T\delta_{ij}\delta_{ab}dt)$~\cite{Gardiner:2010tp}. This yields 
\begin{equation}\label{sSm}
\Delta S_{\rm v} =  \sum_{i=1}^N \int_0^\tau dt \left[ \frac{1}{\gamma T}
|\bm{F}_i|^2 + \frac{1}{m} \bm{\nabla}_{\bm{v}_i} \cdot \bm{F}_i \right]  
    + \frac{1}{\gamma T}\sum_{i=1}^N \int_0^\tau \bm{F}_i \cdot d\bm{W}_i ,
\end{equation}
which is Eq.~(9) of the main text. As explained in the main text, the last
term can be neglected.

The expression for $\Delta S_{\rm v}$ becomes simpler in the polar coordinate. Let $v_i$ and $\theta_i$
are the magnitude and the polar angle of $\bm{v}_i$, respectively. 
The magnitude $\Lambda$ and the polar angle $\psi$ of $\bm{f}$ are given by
$\Lambda e^{i\psi} = \frac{1}{N}\sum_j e^{i\theta_j}$. 
The force $\bm{F}_i = \Gamma (\bm{f} - (\bm{f}\cdot
\hat{\bm v}_i )\hat{\bm v}_i)$ corresponds to the projection of 
$\bm{f}$ in the normal direction of $\bm{v}_i$. Thus, one can write
\begin{equation}
\bm{F}_i = \Gamma \Lambda \sin(\psi-\theta_i) \hat{\bm \theta}_i ,
\end{equation}
where $\hat{\bm \theta}_i$ is the unit vector in the polar angle direction
of $\bm{v}_i$. It is evident that 
$|\bm{F}_i| = \Gamma\Lambda |\sin(\psi-\theta_i)|$. 
The divergence is given by
\begin{equation}
\begin{aligned}
\bm{\nabla}_{\bm{v}_i} \cdot \bm{F}_i &
=\frac{1}{v_i}\frac{\partial}{\partial \theta_i}\Gamma \Lambda 
\sin(\psi-\theta_i)
=\frac{\Gamma}{v_i}\frac{\partial}{\partial
\theta_i}\frac{1}{N}\sum_{j=0}^{N}\sin(\theta_j-\theta_i) \\
& = - \frac{\Gamma}{v_i} \frac{1}{N} \sum_{j\neq i} \cos(\theta_j-\theta_i) 
=\frac{\Gamma}{v_i}\left[\frac{1}{N}-\Lambda\cos(\psi-\theta_i)\right] .
\end{aligned}
\end{equation}
Inserting the magnitude and the divergence of $\bm{F}_i$ into \eqref{sSm}, 
we obtain Eq.~(11) in the main text.

\end{widetext}
\end{document}